\documentclass[10pt, conference]{IEEEtran}

\usepackage{subcaption}   
\usepackage{setspace}

\usepackage{algorithmic} 
\usepackage{algorithm2e}
\renewcommand{\algorithmicrequire}{\textbf{Input:}}
\renewcommand{\algorithmicensure}{\textbf{Output:}}

\usepackage{xspace} 

\newcommand{\ie}{i.e.,\xspace}
\newcommand{\eg}{e.g.,\xspace}

\usepackage{amsthm}
\usepackage{amsfonts}
\usepackage{amssymb}
\usepackage{mathtools}
\usepackage{mathrsfs}

\usepackage{url}

\usepackage{tikz}
\usetikzlibrary{calc}
\usetikzlibrary{positioning}
\usetikzlibrary{shadows}
\usetikzlibrary{fit}
\usetikzlibrary{backgrounds}
\usetikzlibrary{arrows}
\usetikzlibrary{shapes}
\usetikzlibrary{matrix}
\usetikzlibrary{patterns}
\usetikzlibrary{automata}

\tikzstyle{every state}=[draw=black,minimum size=18,inner sep=0,fill=white!65,circle,text=black]

\usepackage{todonotes}

\usepackage{multirow}

\usepackage[figurename=Figure]{caption}

\IEEEoverridecommandlockouts
\IEEEpubid{\makebox[\columnwidth]{978-1-5090-6008-5/17/\$31.00~\copyright~2017 IEEE \hfill} \hspace{\columnsep}\makebox[\columnwidth]{ }}

\begin{document}

\title{A Model for the Analysis of Security Policies \\in Service Function Chains}

\author{
	\IEEEauthorblockN{
										L. Durante, L. Seno, F. Valenza,  A. Valenzano 
	}
	\IEEEauthorblockA{
									National Research Council of Italy (CNR--IEIIT), Corso Duca degli Abruzzi 24, I-10129 Torino, Italy\\
									Emails: \{luca.durante, lucia.seno, fulvio.valenza, adriano.valenzano\}@ieiit.cnr.it
	}
}

\maketitle

\begin{abstract}
	Two emerging architectural paradigms, \ie Software Defined Networking (SDN) and Network Function Virtualization (NFV), enable the deployment and management of Service Function Chains (SFCs). A SFC is an ordered sequence of abstract Service Functions (SFs), \eg firewalls, VPN-gateways, traffic monitors, that packets have to traverse in the route from source to destination. While this appealing solution offers significant advantages in terms of flexibility, it also introduces new challenges such as the correct configuration and ordering of SFs in the chain to satisfy overall security requirements. 
	This paper presents a formal model conceived to enable the verification of correct policy enforcements in SFCs. Software tools based on the model can then be designed to cope with unwanted network behaviors (e.g., security flaws) deriving from incorrect interactions of SFs in the same SFC.

\end{abstract}

\IEEEpeerreviewmaketitle

\section{Introduction}
\label{sec:introduction}

Access to information and services through the Internet has gained fantastic popularity but has also exposed users to cyber-threats and crime. As a recent Verizon report clearly shows~\cite{Verizon2016}, nowadays, attacks carried out through the cyber-space are very tangible menaces. Actually, about 90\% of security breaches discovered in 2015 was aimed at stealing secrets, intellectual properties, personal data and sensitive information. Viruses, malwares, ransomwares, botnets, email spamming and phishing are examples of techniques leveraged by attackers to  pursue their malicious goals.
To mitigate risk, security countermeasures are habitually adopted, such as anti-virus and anti-malware software, personal firewalls, parental control functions and support for virtual private networks (VPN). In the last years, several Internet Service Providers (ISPs) have encouraged this practice and started to offer to their customers security controls, implemented at the ISP premises (e.g. data centers). Of course, providing security services to hundreds of thousands or even millions users is a challenging task which puts into evidence several limitations of traditional network technologies.
Fortunately, Network Function Virtualization (NFV)~\cite{nfv} and Software-Defined Networking (SDN)~\cite{sdn} are two innovative architectural paradigms that can be of significant help in tackling these problems effectively.

It is well known that NFV adopts a virtual infrastructure where network and security functions are implemented
by software applications called Virtual Network Functions (VNF). NFV decouples software and hardware. For instance, VNFs can run in either virtual machines (VMs) or software containers (\ie dockers), hosted on standard high-volume servers (or ad-hoc hardware), so that they can be deployed and removed on-demand.

The SDN paradigm, instead, provides a networking architecture where the control and forwarding planes are kept separated, and control functions are directly programmable. This migration of traffic control, traditionally embedded and hard-wired in individual network devices, to programmable computing equipment enables the abstraction of  the  
communication infrastructure, so that applications and services can be designed, implemented and deployed by considering the network as a logical entity. In summary, SDN introduces unprecedented flexibility in the network, in particular by allowing dynamic fine-grained selections of arbitrary traffic flows that can be (re)routed through different network paths according to the control snap decisions when needed. These features can be leveraged to provide each user with the required security services, as traffic flows concerning different users can be dynamically directed to different sets of devices. 

Service Function Chains (SFCs)~\cite{RFC7665} are strictly related to the adoption of NFV and SDN. A SFC can be considered as an ordered sequence of Service Functions (SFs) such as NAT, firewalls, QoS and so on, so that packets have to traverse a sequence of cascaded services in their route from source to destination.
%
Of course, this scenario introduces new challenges involving the correct design, configuration and ordering of service functions~\cite{RFC7498}, for instance to prevent conflicts and/or inconsistencies, in particular when different subnetworks belong to different administration domains. 

At present, the configuration of most service functions relies on low-level specific parameters, whose values have to be selected and set manually. In practice, this implies a typical configuration approach by trials and errors. Frequently, when a misconfiguration is  detected, administrators try to correct errors by introducing ad-hoc rules and repeat this process until  no more anomalies are observed. This technique, beside being cumbersome and  time-consuming, lacks a comprehensive view of the network behavior and, as such, is error-prone and can make the network significantly hard to maintain.
Clearly, the independent configuration of functions is not suitable to grant security in the whole network, as also recognized by the working group on the Interface to Network Security Functions (I2NSF) in one of their published reports~\cite{I2NSF},  because in distributed environments also interactions between different NSFs  have to be carefully considered.

Because of the size and complexity of modern networks, preventing conflicts in SFC configurations  is a very hard task, especially without the aid of suitable automated software tools. Practical solutions, however, have to be based on ad-hoc formal descriptions and sound theoretical foundations.

In this paper we propose a novel model conceived to check the security policy configuration in SFCs consisting of heterogeneous functions. The model is able to describe the typical actions carried out to process network traffic (\eg packet header modification, payload encryption, tunneling) by taking into account both individual service functions and their sequential combination.

The paper is structured as follows. Section~\ref{sec:related_works} briefly recalls some works appeared in the literature, which focus on the analysis of policy conflicts, service function modeling and policy verification. Section~\ref{sec:approch} presents the proposed approach, while Section~\ref{sec:model} deals with the formal model for traffic transformations carried out by service functions. Section~\ref{sec:verification} gives some hints about the policy verification approach. Finally, Section~\ref{sec:conclusions} concludes the paper.


\section{Related works}
\label{sec:related_works}
In general, policy conflicts and their analysis have been tackled in a number of literature papers. 
In the area of network security, several studies have dealt with firewall and IPSec-VPN policies, while
little work has been done on the verification of  policies in SFCs.

An anomaly analysis of filtering policies in distributed scenarios was first presented in \cite{Al-Shaer2005}. The authors introduced a classification scheme for packet filtering rule relations and defined four types of intra-policy rule anomalies (shadowing, correlation, generalization and redundancy), as well as five types of intra-policy anomalies (shadowing, spuriousness, redundancy, correlation and irrelevance). In \cite{Liu2005} Liu et al. focused on the detection and removal of redundant rules. 
They proposed a classification of redundant rules including (never matched) upward and (matched) downward elements, the latter enforcing the same actions as other lower priority rules.
IPSec and VPN anomalies were studied in \cite{Fu2001}, \cite{Al-Shaer2005IPSec} and \cite{Basile2014,basile2016}. In \cite{Fu2001} the authors presented an approach to detect IPSec policy anomalies. Their analysis was based on a set of policy implementations described in a high-level language and anomalies were identified by checking the implementations against the expected behavior.
In \cite{Al-Shaer2005IPSec}  Hamed and Al-Shaer formalized the classification scheme of \cite{Fu2001}. Their model included both the encryption and packet filter capabilities of IPSec. In particular, the authors identified two new IPSec errors (overlapping-session and multi-transform anomalies), that are of interest for both the inter and intra-policy analyses.
In \cite{Basile2014} Basile et al. presented a new classification for inconsistencies in policies for communication protection. The proposed model enables the detection of a number of problems arising from complex interactions involving protection protocols at different layers of the communication stack, security properties, end-to-end communication channels, VPNs and remote-access communications.
All these works focus on the policy abstract world and do not take into account the actual policy implementation in real systems.
%
%
%
%

The verification of the whole system correct behavior has been another hot topic since the introduction of the SDN and NFV paradigms, as SDN enables multiple applications/users to (re)program the same physical network. The discovery of network-wide invariant violations (i.e. presence of forwarding loops, black holes and so on) is a crucial issue, in particular when it has to be carried out in real-time and without affecting the system performance. 
Modern SDN controllers are able to manage around 30K new flows installs per second while maintaining a sub-10 ms flow install time \cite{Tavakoli2009}. In doing so, however, they have to check for updates and analyze the  invariants before changes affect the network, raising alarms when needed, and preventing errors by blocking inappropriate modifications. 

From this point of view, for instance, VeriFlow \cite{Khurshid2012} makes use of graph search techniques to verify network-wide invariants and manages dynamic changes in real-time. The tool can discover faulty rules issued by SDN applications and optionally prevent them from reaching the network causing anomalous behavior.
ConfigChecker \cite{Al-shaer2009}, instead, converts the (configuration and forwarding) rules into boolean expressions that are then checked against network invariants.  However, VeriFlow and ConfigChecker are first generation of SDN verification tools and, thus, they overlook the presence of stateful functions in the network.
Also SFC-Checker~\cite{Tschaen2016} is a framework able to perform correctness verification of packet forwarding in SFCs, but it does not consider security functionalities.

Finally, isolation properties have been studied in \cite{Panda2014, Spinoso2015} for networks that include dynamic data paths, by means of model checking techniques. In this case, the notion of verification is first extended to cope with systems containing dynamic paths, before checking invariants such as connectivity or isolation.

\section{Approach overview}
\label{sec:approch}
%
According to the RFC-7665 document~\cite{RFC7665}, an SFC system consists of three main components:
\begin{itemize}
	\item the \textit{Service Function Path} (SFP), which specifies the sequence of SFs a packet should go through;	
	\item the \textit{Classificator and Encapsulator} (SFCE), which is responsible for classifying and encapsulating packets (\ie SFP selection);
	\item the \textit{Service Function Forwarder} (SFF), which manages the SFP and routes packets between SFs according to information encoded in the encapsulation.	
\end{itemize}
SFC implementations can take advantage of the SDN paradigm as in the case of the standard OpenFlow  communication interface~\cite{Openflow2014}, which is currently the most popular solution adopted in SDN architectures. In this framework, \textit{OpenFlow  Flow Entries} are used to assign each traffic flow a corresponding SFP, the \textit{OpenFlow controller} carries out the SFCE activities according to the OpenFlow Flow Entries, while \textit{OpenFlow switches} behave as the SFF.

Assuming a generic SFC implementation based on OpenFlow, our verification process is based on the following steps:
\begin{enumerate}
	\item OpenFlow Flow Entries are collected and applied to obtain the set of SFCs and corresponding SFPs;
	\item for each SFC, appropriate verification policies (\ie security requirements to be enforced by each SFC on its traffic flows) are defined;	
	\item low-level configuration policies shaping the SFs of each SFC are retrieved from the SF implementation;
	\item the formal model is populated with the verification and low-level configuration policies;
	\item transformations performed by SFs are computed for any traffic flow, and results are checked against the verification policies.	
\end{enumerate}
%

\begin{figure}[t]
	\centering
	\includegraphics[width=\linewidth]{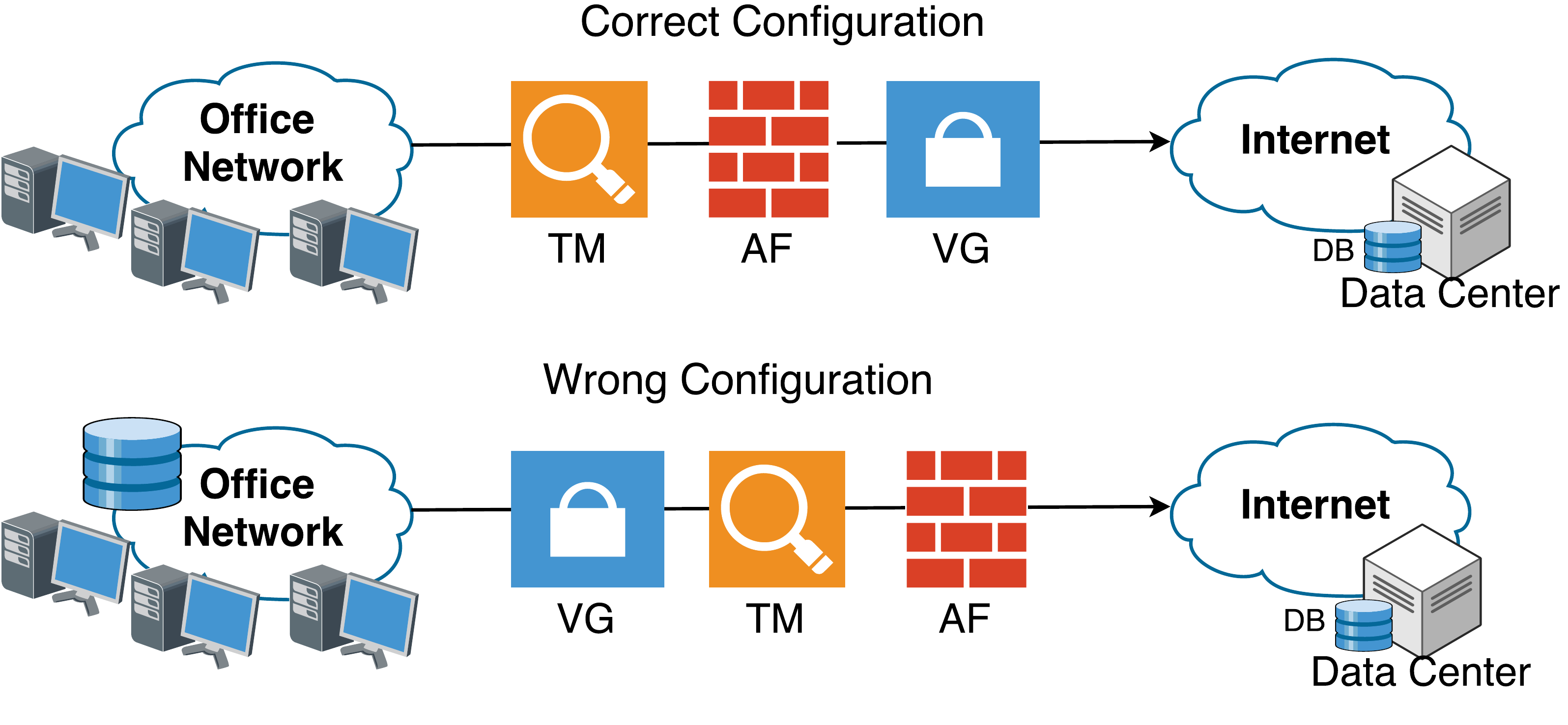}
	\caption{Simple SFCs}
	\label{fig:sfcexample}
\end{figure}

As an example, let us consider the very simple scenario shown in Figure~\ref{fig:sfcexample}, where users in the office network access the Internet through a short SFC consisting of a Traffic Monitor (TM), an Application Firewall (AF) and a VPN Gateway (VG). A company database (DB) is hosted in a remote Data Center (DC) and three main requirements have to be satisfied: 
\begin{enumerate}
	\item all encrypted private traffic from office users must be dropped (for instance to prevent the undetected disclosure of company confidential information);
	\item all traffic to DC must be encrypted;
	\item connections to DB have to be monitored. 
\end{enumerate}
To cope with these constraints, network administrators can  configure the SFs so that:
\begin{itemize}
	\item TM implements the \textit{``count connections to DB''} policy;
	\item AF implements the \textit{``drop all encrypted outgoing traffic''} policy;
	\item VG implements the \textit{``encrypt all traffic to the DC''} policy.
\end{itemize}
This simple configuration is correct if and only if the order of functions in the SFC is \textit{TM-AF-VG} (``\textit{Correct Configuration}" in Figure~\ref{fig:sfcexample}). For instance, ordering the SFC sequence as \textit{VG-TM-AF} would lead to a situation where VG encrypts all traffic to DC, but TM cannot count the number of connections and AF drops all traffic sent to DC.

The above example will be employed in section \ref{sec:verification} to clarify our model and the verification process we propose.

\section{Model description}
\label{sec:model}
The formal model introduced in this section is able to describe a single SFC, the traffic passing through it and the policies the SFC is intended to enforce (verification policies). If many SFCs are implemented by an ISP in the same network infrastructure, the model and, consequently, the verification process may be employed for each of them separately.

Moreover, we suppose that ISPs use a number of Service Functions to offer a variety of network operations, such as traffic filtering, monitoring, rewriting etc. For this reason, we define a modeling approach that is flexible enough to include a rich set of functions. Examples of common Service Functions the model is able to describe are Packet Filters, Application Firewalls, NATs, VPN gateways, DPIs and IDSs.

\subsection{Service Function Chains and Service Functions}
A service function chain $SFC$ is modeled as an ordered finite set of service functions
$$
SFC=[SF_1, SF_2, \dots, SF_{n_{SFC}}]
$$
A service function $SF$ is a pair
$
SF=(P,S)
$,
where $P$ represents the policy implementing $SF$ and consists, basically, of an ordered finite set of rules, while $S$ is the current state of $SF$, modeling the values assumed by the set of variables stored in the internal state table of the service function, if any. We use notations $SF.P$ and $SF.S$ to refer, respectively, to the policy and internal state table of $SF$. The formal description of $P$ and $S$ is provided further on in this section. 

\subsection{Traffic} 
A packet $p$ is a finite set of pairs, each one consisting of a network field and its value. Formally
$$
p=\{(n_1,v_1), (n_2,v_2),\dots, (n_{n_p},v_{n_p})\}
$$
A network field may represent, for instance, the source or destination address in an IP packet, the port number in a UDP packet, the ACK in a TCP packet or the MIME type in an HTTP packet. We suppose all possible network fields $n$ to be known and collected in a finite set $\mathcal{N}$. Moreover, each network field $n \in \mathcal{N}$ is characterized by its own length, corresponding to the field size in number of bits. 
For sake of conciseness, we also use notation $p(n_i)$ to mean the value assumed by network field $n_i\in\mathcal{N}$ in packet $p$ (\ie $p(n_i)=v_i, \ i=1,\ldots, n_p$).

Under these hypotheses, the packet space $\mathcal{P}$ (\ie the set of all possible packets) is the set of all finite subsets of $\mathcal{N}$ (consistent from real packet description prospective), associated to all  possible values they can assume. Clearly, a real packet does not contain all network fields in $\mathcal{N}$ but only a limited subset of them. Moreover, we extend $\mathcal{P}$ to include the null packet $\emptyset$,  (\ie a packet consisting of no network fields), to model dropped elements.
Then, a traffic ${T}$ can be represented as a finite ordered set of packets
$$
T=[p_1, p_2,\dots, p_{n_T}],\quad p_i\in\mathcal{P}
$$

\subsection{State of a Service Function} 
A function $SF$ can be stateful, \ie it records its current state in an internal table consisting of a set of variables (state fields). State fields are updated over time, depending on the input traffic characteristics,  and affect the traffic transformation performed by $SF$. Stateful firewalls, for example, perform packet filtering by keeping track of the connection status. Similarly, actions performed by intrusion detection systems (IDSs) and packet/connection counters may depend on the maximum number of HTTP connections opened from the same IP address, the number of source IP addresses for the same user, or the current session duration time.

We model the state $S$ of a service function $SF$ as a finite set of pairs, each consisting of a state field and its value. Formally
$$
S=\{(s_1,v_1), (s_2,v_2), \dots, (s_{n_S},v_{n_S})\}
$$
The set of fields $s_i$ composing the SF state may vary consistently among different service functions. 
The finite set of all possible state fields is known:  $\mathcal{S} = \{s\}$. We use notation $SF.S(s_i)$ to refer to the value assumed by state field $s_j\in \mathcal{S}$ in state $S$ of $SF$ (\ie $SF.S(s_i)=v_i, \ i=1,\ldots, n_S$) and $S=\emptyset$ when a service function is stateless.

Finally $SFC.S^{\cup}$ is the global state of a service function chain $SFC=[SF_1, SF_2, \dots, SF_{n_{SFC}}]$, \ie 
$$
\mathcal{S}^{\cup}=\bigcup SF_{i}.S, \quad i=1, \ldots, n_{SFC}
$$

\subsection{Policy Implemented by SF}
A service function $SF$ processes an input packet by performing some actions and releasing a modified version of the packet itself. Typically, it is implemented through a set of rules, which represents the SF policy. An SF policy $P$ is a triple
$$
P=\left(R,\mathfrak{R}, a_{d}\right)
$$
where
\begin{itemize}
	\item $R=[r_1, r_2, \ldots, r_{n_R}]$ is the ordered finite set of rules implementing the policy;
	\item $\mathfrak{R}$ is the resolution strategy employed by SF to decide the action to be applied when the input packet matches more than one rule (\eg a common resolution strategy is the first matching rule (FMR), where the rule with the highest priority is applied);
	\item $a_{d}$ is the default action, which is applied when no rule is matched.
\end{itemize}

According to the RFC-3198~\cite{RFC3198} document, each rule $r_i$ is modeled as a set of conditions and an ordered finite set of actions, \ie
$$
r_i=(C,A)=(\{c_1,c_2, \dots, c_{n_C}\}, [a_1, a_2, \dots, a_{n_A}])
$$
$C$ is the set of prerequisites (conditions) that have to be satisfied for the sequence of actions $A$ to be orderly performed. Note that, conditions and actions may involve both the input packet and the service function state. In detail, any condition $c \in C$ is a triple
$$
c=(f,\rho,v)
$$
where
\begin{itemize}
	\item $f$ is either a network or a state field (\ie $f \in \mathcal{N} \ \vee \ f \in \mathcal{S}$)
	\item $\rho$ is a relation (\eg $=,>,\subset,\nsupseteq,\in$);
	\item $v$ is the value of either a network or a state field (\eg 1.1.1.1, 8080).
\end{itemize}
We also use the following notations
\begin{itemize}
	\item $f=\mathcal{N}$ and $f=\mathcal{S}$ mean any network or state fields, respectively;
	\item $\rho=*$ is any possible relation;
	\item $v=*$ is any possible value among those admissible for field(s) $f$ (in some cases, we also use hybrid notations, \eg 1.1.1.* is a set of values for a source IP address);
	\item in this paper, for sake of simplicity, $v=\bullet$ means encrypted value(s)
\end{itemize}

Under these hypotheses, a packet $p$ processed by $SF$ with current state $SF.S$, satisfies condition $c_i \in SF.P.C$ if at least one of the following two predicates holds true
\begin{eqnarray*}
&\-&f_i \in \mathcal{N} \ \wedge \ (p(f_i) \; \rho_i \; v_i \ \vee \ \nexists  \ p(f_i)) \\
&\-&f_i \in \mathcal{S} \ \wedge \ (SF.S(f_i) \; \rho_i \; v_i \ \vee \ \nexists \ SF.S(f_i)) 
\end{eqnarray*}
For example, condition $c_1=(f_1,\rho_1,v_1)$, where $f_1= ip\_dst$, $\rho_1= \ \subset$ and $v_1 = {1.1.1.*}$, is satisfied by all packets with destination IP address in the range $1.1.1.0/24$. 


All possible actions a service function can carry out are known and belong to a finite set $\mathcal{A}=\{a\}$. Any action $a\in \mathcal{A}$ is a function that modifies either the input packet, the SF state or both, possibly according to a set of input parameters $\{e\}\subseteq \mathcal{E}$. Formally
$$
a: (\mathcal{P}, \mathcal{S}, \mathcal{E}) \rightarrow (\mathcal{P}, \mathcal{S}) 
$$ 
Some examples of common actions that can be described according to the proposed model are the following:

\begin{itemize}
	
	\item $a_{\text{ALLOW}}$: this action leaves all packet network fields and SF state fields unchanged, \ie formally $a_{\text{ALLOW}}(p,S)=(p,S)$;
	
	\item $a_{\text{DENY}}$: this action blocks a packet: $a_{\text{DENY}}(p, S)=(\emptyset, S)$, \ie it transforms any input packet into a null packet while leaving the state unchanged;
	
	\item $a_{\text{MOD\_NF}}$: many actions modify only one (some) specific network field(s) within the input packet; in these cases the function parameters are the pair(s) consisting of affected network field and the value it assumes after the action is performed (for instance, a NAT action has as input parameters the source address field and the IP address of the NAT, \ie $a_{\text{MOD\_NF}}(p, S, \{(ip\_src, 1.1.1.1)\})=(p(ip\_src)=1.1.1.1, S)$;
	
	\item $a_{\text{MOD\_SF}}$: actions performed by service functions like network monitoring, logging and counting often only affect specific state field(s) (input packets are only read by these functions); in these cases, parameters are the pair(s) consisting of affected state field and its new value after the action is performed;
	
	\item $a_{\text{ENCAPSULATE}}$: the encapsulation of an input packet involves adding network fields to the packet and, as such, takes as input(s) the network field(s) that need to be added and the related value(s). 
	
	\item $a_{\text{ENCRYPT}}$: the action of encrypting a packet, or a (set of) field(s) in the packet, is basically a specific case of $a_{\text{MOD\_NF}}$ action; we assume the corresponding function to take as an input the field(s) to be encrypted and the kind of encryption to be performed.  As an example, the IPSec encryption of the layer 4 payload ($P\!L_4$) is described by $a_{\text{ENCRYPT}}(p, S, \{(P\!L_4,  \text{ike=aes256-sha512-modp4096},$\\$ \text{esp=aes256-sha512-modp4096})\})= (p.P\!L_4=\bullet, S)$ .
\end{itemize}

The rule definition allows to combine many actions that are sequentially performed any time all conditions of the rule are satisfied. Given a policy $P$ we can define the transformation $\mathfrak{T}_P$ associated to $P$ which describes how a specific pair, consisting of an input packet and an SF state, is modified by $P$. Formally, $\mathfrak{T}_P$ is a function
$$
\mathfrak{T}_P:(\mathcal{P},\mathcal{S})\rightarrow(\mathcal{P},\mathcal{S})
$$ 
$$
\mathfrak{T}_P(p,S)=   a_1 \circ a_2 \circ... \circ a_{n_A} (p,S) 
$$
where $r=(C,A)=(\{c_1,c_2, \dots, c_{n_{C}}\}, [a_1, a_2, \dots, a_{n_{A}}]) \in P.R$ is the rule matched by the input packet according to $\mathfrak{R}$ (note that symbol $\circ$ indicates function composition). 
Since a network traffic $T_i$ is defined as a sequence of packets we extend $\mathfrak{T}_P$ so that, when applied to a pair $(T_i,S_i)$, it returns
$$
\mathfrak{T}_P({T}_i,{S}_i)=({T}_f,{S}_f)=(\circ \mathfrak{T}_P(p \in T_i, S_i))
$$ 
that is, the sequence of packets obtained by applying $\mathfrak{T}_P$ to the sequence of input packets $p\in T_i$ and the initial state $S_i$.

\subsection{Verification Policies}
The set of verification policies (\ie, the policies that are supposed to be implemented by the service function chain) is
$$
V=\{v_1, v_2, \dots, v_{n_V}\}
$$
Each verification policy $v\in V$ is a quadruple specifying an input network traffic ($T_i$), supposed to pass through the SFC, an initial global state (${S}^{\cup}_i$), and the corresponding expected output network traffic ($T_e$) and final global state (${S}^{\cup}_e$). Formally
$$
v=(T_i, {S}^{\cup}_i, T_e, {S}^{\cup}_e )
$$
Note that, when we consider stateless service functions $v=(T_i, \emptyset, T_e, \emptyset)$. We adopt notation $v.T_i$,  $v.T_e$,  $v.{S}^{\cup}_i$ and  $v.{S}^{\cup}_e$ to refer to, respectively, initial and expected output network traffic and initial and expected final global state defined by $v$.

\section{Verification}
\label{sec:verification}
Given a SFC, the traffic passing through it and the verification policies to be enforced by the SFC, the verification process can be carried out following the steps shown in Algorithm~\ref{Alg}. Actually, the algorithm takes the description $SFC$ and verification policies $V$ as inputs, and returns sets $P_{true}$ and $P_{false}$ containing the correctly enforced and disregarded policies, respectively. \ 

To clarify how the model can be used to describe real SFCs and how the verification process is then carried out, we present a simple example inspired by the SFCs shown in Figure~\ref{fig:sfcexample}. 

Firstly, models for the SFs have to be specified, by describing both the SF initial states and the policies they implement. 
The traffic monitor ($SF_{tm}$) is described as $SF_{tm} = (P_{tm}, S_{tm}^0) = ((C_{tm}, A_{tm}), S_{tm}^0)$, where $C_{tm}$ is verified by all packets whose destination IP address is DB. Correspondingly, action $A_{tm}$ increments the number of opened connections to the database. The initial state of $SF_{tm}$ ($S_{tm}^0$) keeps track of the number of database connections at the start-up time. Formally
\begin{gather*} 
	S_{tm}^0=\{(con\_db,0)\}, \;	C_{tm}=\{(ip\_dst,=,IP\_DB)\}, \\
	A_{tm}=[a_{\text {MOD\_SF}}(\{(con\_db, +1)\}]
\end{gather*}  
The application layer firewall is modeled as a stateless service function, $SF_{af} = ((C_{af},A_{af}),\emptyset)$, whose condition $C_{af}$ catches all packets with encrypted level 4 payload so that the associated action $A_{af}$ can discard them
\begin{gather*} 
	C_{af}=\{PL_4,=,\bullet)\},\; A_{af}=[a_{\text {DENY}}] 
\end{gather*} 
Finally, the VPN gateway is a stateless service function that encrypts packets to the data center, \ie $SF_{vg} = ((C_{vg},A_{vg}),\emptyset)$, where
\begin{gather*} 
	C_{vg}=\{(ip\_dst, \in, NET\_DC)\},\\
	A_{vg}=[a_{\text {ENCAPSULATE}}(\{\ldots\}), a_{\text {ENCRYPT}}(\{\ldots\})] 
\end{gather*}  \
We can now define the set of verification policies as $V=\{v_1,v_2\}$, where $v_1 = (T_i^1, {S}^{\cup 1}_i, T_e^1,{S}^{\cup 1}_e )$ means that connections to the DB service (e.g., GET methods in HTTP packets) have to be encrypted and the connection counter increased by one unit at the same time. Similarly,  $v_2 = (T_i^2, \emptyset, T_e^2,\emptyset)$ means that encrypted packets from the office network must be dropped. Formally
\begin{gather*} 
T_i^1=[\{(ip\_dst, IP\_DB), (http\_metod, GET), \ldots\}]\\
T_e^1=[\{(ip\_src, IP\_GW),(PL_4,\bullet), \ldots\}]\\
{S}^{\cup 1}_i=\{(con\_db,0)\},  \;\; {S}^{\cup 1}_e=\{(con\_db,1)\}\\
T_i^2=[\{(ip\_src, IP\_employee), (PL_4,\bullet), \ldots \}], \;
T_e^2=\emptyset
\end{gather*} 
Let us assume that SFC is configured as $SFC=[SF_{vg}, SF_{tm}, SF_{af}]$ (bottom SFC in Figure~\ref{fig:sfcexample}). By applying Algorithm \ref{Alg}, the verification process detects anomalies. Specifically, policy $v_1$ is not correctly enforced as the VPN gateway encrypts the incoming packets (since they are sent to database). As expected, the traffic monitor cannot update the connection counter (it is not able to read the packet IP destination address which is encrypted) and, finally, the application firewall drops the packets because of their encryption. Formally
\begin{gather*} 
	(T_1,S_1)=\mathfrak{T}_{vg}|_{P_{vg}}(T_i^1, {S}^{\cup 1}_i)=(T_e^1, {S}^{\cup 1}_i)\\
	(T_2,S_2)=\mathfrak{T}_{tm}|_{P_{tm}}(T_1, {S}_1)=(T_e^1, {S}^{\cup 1}_i)\\
	(T_3,S_3)=\mathfrak{T}_{af}|_{P_{af}}(T_2, {S}_2)=\mathbf{(\emptyset, {S}^{\cup 1}_i)\neq(T^1_e, {S}^{\cup 1}_e)}
\end{gather*}  
Conversely, policy $v_2$ appears to be correctly enforced as the application firewall drops any encrypted packet
\begin{gather*} 
	(T_1)=\mathfrak{T}_{vg}|_{P_{vg}}(T_i^1)=(T_e^1)\\
	(T_2)=\mathfrak{T}_{tm}|_{P_{tm}}(T_1)=(T_e^1)\\
	(T_3)=\mathfrak{T}_{af}|_{P_{af}}(T_2)=\mathbf{(T_e^2)}
\end{gather*}  
When the SFC is configured as in the upper part of Figure~\ref{fig:sfcexample}, instead, both policies $v_1$ and $v_2$ are checked as correct. Indeed, the verification process for policy $v_1$ shows that the traffic monitor increments the number of connections for packets sent to the database, the application firewall forwards (cleartext) packets and the VPN gateway performs encryption as requested by $v_1$. Formally
\begin{gather*} 
	(T_1,S_1)=\mathfrak{T}_{tm}|_{P_{tm}}(T_i^1, {S}^{\cup 1}_i)=(T_i^1, {S}^{\cup 1}_e)\\ 
	(T_2,S_2)=\mathfrak{T}_{af}|_{P_{af}}(T_1, {S}_1)=(T_i^1, {S}^{\cup 1}_e)\\ 
	(T_3,S_3)=\mathfrak{T}_{vg}|_{P_{vg}}(T_2, {S}_2)=\mathbf{(T_e^1, {S}^{\cup 1}_e)} 
\end{gather*}  
Policy $v_2$ is also correctly  implemented as (already) encrypted packets from the office network are dropped by the application firewall
\begin{gather*} 
	(T_1)=\mathfrak{T}_{tm}|_{P_{tm}}(T_i^1)=(T_i^1)\\
	(T_2)=\mathfrak{T}_{af}|_{P_{af}}(T_1)=(T_e^2)\\
	(T_3)=\mathfrak{T}_{vg}|_{P_{vg}}(T_2)=\mathbf{(T_e^2)}
\end{gather*}  
\begin{algorithm}[t]
	\algorithmicrequire{ SFC ${SFC}$, verification policy set $V$ }\\
	\algorithmicensure{ verification process results $P_{true}$, $P_{false}$}\\
	\ForAll{verification policy $v \in V$ }{
		$T_f=v.T_i$, ${S}^\cup_f=v.{S}^\cup_i$\\
		\ForAll{service function $SF \in SFC$}{
			${P}= SF.P$\\
			$(T,{S})= \mathfrak{T}_{P}(T_f,{S}^\cup_f) $\\
			$T_f={T}$\\
			${S}^\cup_f={S}^\cup_f\cup {S}$\\
		}
		\eIf{$T_f==v.T_e \ \wedge \ {S}^\cup_f==v.{S}^\cup_e$}{
			$P_{true}=P_{true}\cup \{v\}$
		}{
			$P_{false}=P_{false} \cup \{v\}$
		}
	}
	\caption{Algorithm for security policy verification}
	\label{Alg}
\end{algorithm}
%


\section{Conclusions}
\label{sec:conclusions}
A formal model for the analysis of security policies in SFCs has been presented in this paper. The model is able to describe, in a simple way, a broad set of SFs, such as those carried out by firewalls, NAT/NAPT devices, traffic monitors and encryption equipment. The main goal of the model is to pave the way to the automatic verification of correct policy implementation in SFCs by means of suitable automatic software tools.  

Future research activities will be oriented to refining the model and making it suitable to deal with typical issues such as policy reachability and reconciliation. This step, however, involve a deeper knowledge of the network to ensure that the best policies are selected and to solve policy conflicts. 


We also plan to develop a verification tool, which that will be able to take care of the verification process in an open-source NFV architecture, which may may be employed over multiple underlaying physical domains (e.g. federated networks). For integrating our approach in an NFV environment several challenges need to be addressed, such as providing the NFV Management and Orchestrator with the necessary information to deal with security verification.



\end{document}